\documentclass[smallextended]{svjour3}       
\smartqed  
\usepackage{graphicx}
\usepackage{amsmath,amssymb}
\usepackage[margin=1in]{geometry}
\usepackage{graphicx}
\usepackage{subfigure}
\usepackage{multirow}
\usepackage{array}
\usepackage{makecell}
\usepackage{float}
\usepackage{booktabs}
\usepackage{xcolor}
\usepackage{mwe}
\usepackage{lineno}
\usepackage{epsfig}
\usepackage{xr}
\usepackage{natbib}

\newcommand{\E}{\mathop{\rm E}\nolimits} 
\newcommand{\Var}{\mathop{\rm Var}\nolimits} 
\newcommand{\sign}{\mathop{\rm sign}\nolimits} 
\journalname{Computational Statistics}
\begin{document}

\title{Smoothed quantile regression for censored residual life}

\author{Kyu Hyun Kim \and Daniel J. Caplan \and
	Sangwook Kang
}


\institute{Kyu Hyun Kim
	\at Yonsei University, 50 Yonsei-ro, Seodaemun-gu, Seoul, Korea \\
	Tel.: +82-2-2123-2535\\
	Fax: +82-2-2123-8638\\
	\email{kyuhyunkim07@yonsei.ac.kr} 
	\and
	Daniel J. Caplan
	\at Department of Preventive and Community Dentistry, College of Dentistry, University of Iowa, 
	Iowa City, IA, USA \\ 
	Tel.: +1-319-335-7184 \\
	Fax: +1-319-335-7187\\
	\email{dan-caplan@uiowa.edu}
	\and
	Sangwook Kang
	\at Yonsei University,50 Yonsei-ro, Seodaemun-gu, Seoul, Korea \\ 
	Tel.: +82-2-2123-2538\\
	Fax: +82-2-2123-8638\\
	\email{kanggi1@yonsei.ac.kr}
}

\date{Received:  / Accepted: }

\maketitle

\begin{abstract}
We consider a regression modeling of the quantiles of residual life, remaining lifetime at a specific time. We propose a smoothed induced version of the existing non-smooth estimating equations approaches for estimating regression parameters. The proposed estimating equations are smooth in regression parameters, so solutions can be readily obtained via standard numerical algorithms. Moreover, the smoothness in the proposed estimating equations enables one to obtain a robust sandwich-type covariance estimator of regression estimators aided by an efficient resampling method. To handle data subject to right censoring, the inverse probability of censoring weight are used as weights. The consistency and asymptotic normality of the proposed estimator are established. Extensive simulation studies are conducted to validate the proposed estimator's performance in various finite samples settings. We apply the proposed method to dental study data evaluating the longevity of dental restorations.
\end{abstract}

\section{Introduction} \label{sec:intro}
Remaining lifetimes at a specific time are frequently of interest in many clinical and epidemiological studies. In contrast to the conventional failure time, which is defined as the time elapsed from the baseline until an event occurs, residual life can be defined at any time $t$ after the baseline provided the subject has not experienced the event of interest by $t$. Because survival data is frequently collected through a series of follow-up visits following an initial baseline visit, modeling residual life at a specific visiting time $t$ could provide more dynamic and meaningful information. In a dental study, for example, the longevity of a tooth treated with a restoration procedure may be of interest. Subjects who have had a tooth restored typically return to the clinic on a regular basis for a check-up. It would be very interesting to assess the effects of factors that might be related to the residual life of the treated tooth and predict its longevity at each visit. In this case, your main concern is, for example, how long will my restored tooth last if it is still alive one year after restoration?

To investigate the effects of various factors on remaining lifetimes, regression modeling have typically been on the mean and quantiles of residual life. Modeling the mean residual life has mostly focused on a proportional mean residual life model, a counterpart of Cox proportional hazard model \citep{maguluri1994estimation,oakes2003inference,chen2005semiparametric,bai2016semiparametric}. Statistical methods also have been proposed under alternative models including an additive mean residual life model \citep{chen2006linear,chen2007additive, zhang2010goodness, mansourvar2016additive, cai2017additive}, proportional scaled mean residual life model \citep{liu2008regression} which can be considered as the accelerated failure time (AFT) model, and semiparametric transformation models \citep{sun2009class,sun2012mean, cheng2014combined}. Although a mean has been a popular quantity representing the central tendency in data, it might not be a suitable summary measure for survival times that are typically skewed, possibly with a heavy tail and extreme observations. Moreover, the identifiability could be an issue for the mean residual life when the follow-up time is not long enough \citep{li2016quantile}. In such cases, a median, a special case of quantiles, could serve as a nice alternative; the median is a less sensitive measure to outliers and thus offer a more meaningful summary for skewed survival data \citep{sit2017survival}.

Quantile regression models, originally proposed for modeling a continuous response \citep{koenker1978regression}, have been adapted to modeling failure time data without considering censoring \citep{jung1996quasi,portnoy1997gaussian,wei2006quantile,whang2006smoothed} and expanded to accommodating censored failure time data \citep{ying1995survival,bang2002median,portnoy2003censored,peng2008survival,wang2009locally,huang2010quantile,portnoy2010asymptotics}. A recent work of \citet{peng2021quantile} provides a comprehensive review of statistical methods developed for fitting quantile regression models with various types of survival data. For residual life, \citet{jung2009regression} considered a semiparametric regression model and proposed estimating equations approach. \citet{kim2012censored} proposed an alternative estimating equations approach, for which an empirical likelihood approach is taken. \citet{li2016quantile} considered a more general setting that allows repeated measurements of covariates. \citet{bai2019general} proposed a general class of semiparametric quantile residual life models for length-biased right-censored data. They proposed an estimating equations approach that employs the inverse probability of censoring weighted (IPCW) principle to handle right-censored observations. Asymptotic properties of the proposed estimators were rigorously established in all these approaches.

It is important to note that all of the estimating equations considered in these proposed works have non-smooth regression parameters. As a result, standard numerical algorithms such as Newton-Raphson cannot be directly applied in calculating the proposed estimators, and solutions may not be uniquely defined. Nevertheless, recent developments in computing algorithms have alleviated these issues substantially. Efficient and reliable calculation of regression parameters estimates have been shown to be feasible. Regression coefficients estimators were calculated via a grid search method \citep{jung2009regression} or $L_1$-minimization algorithm based on the linear programming technique \citep{kim2012censored,li2016quantile}. Despite these progresses in point estimation, variance estimation still could be problematic. Because the proposed estimating functions are not differentiable by regression parameters, a well-known robust sandwich-type estimator cannot be calculated directly. Furthermore, a direct estimation of variance involve nonparametric estimation of unspecified conditional error density, which could be computationally unstable. For these reasons, either a direct estimation of variance was avoided \citep{jung2009regression,kim2012censored} or a computationally intensive multiplier bootstrap method was used, which required solving perturbed estimating equations multiple times. One approach to this problem is to use an induced smoothing procedure, which converts non-smooth estimating equations into smooth ones.

The induced smoothing approach \citep{brown2005standard} has been frequently employed in survival analysis especially for the rank-based approach in fitting semiparametric AFT models \citep{brown2007induced,johnson2009induced,fu2010rank,pang2012variance,chiou2014fast,chiou2014fitting,chiou2015rank,chiou2015semiparametric,Kang:fitt:2016} and quantile regression models \citep{choi2018smoothed}. Based on the asymptotic normality of the non-smooth estimator, the non-smooth estimating functions are smoothed by averaging out the random perturbation generated by adding a scaled mean-zero normal random variable to the regression parameters. Because the resulting estimating functions have smooth regression parameters, they can be easily applied using standard numerical algorithms such as Newton-Raphson. Furthermore, due to the smoothness, the estimation of standard errors is straightforward by using the robust sandwich-type estimator. This induced smoothing approach has not yet been applied in the context of quantile residual life regression, to our knowledge. Thus, we propose to apply the induced smoothing technique to the fitting of a semi-parametric quantile residual life regression model.

The rest of the article is organized as follows. In Section~\ref{sec:model}, a semiparametric regression model for quantiles of residual life and the proposed estimation methods in the presence of right censoring are provided. In Section~\ref{sec:comp}, two computing algorithms are introduced and proposed estimator's asymptotic properties are established in Section~\ref{sec:asymp}. Finite sample properties of the proposed estimators are investigated by simulation experiments in Section~\ref{sec:sim} and the proposed methods are illustrated with a dental restoration longevity study data for older adults \citep{caplan2019dental} in Section~\ref{sec:realdata}. Concluding remarks are provided with some discussions in Section~\ref{sec:discussion}.

\section{Model and Estimation} \label{sec:model}
\subsection{Semiparametric quantile regression model} \label{sec:quantile}

Suppose $T$ and $C$ denote the potential failure time and censoring time, respectively. In the presence of right censorship, we observe $Z = \min(T,C)$, which $T$ and $C$ are independent. Let $\delta = I[T \leq C]$ be the failure indicator where $I[\cdot]$ is an indicator function. Then, with a $p \times 1$ vector of covariate $X$, we observe $n$ independent copies of $(Z, \delta, X)$, $(Z_i, \delta_i, X_i), i = 1, \ldots, n$, where $n$ and $i$ denote the sample size and subject, respectively. At time $t_0$, the $\tau$-th quantile of the residual life is defined as $\theta_{\tau}(t_0)$ that satisfies $P(T_i - t_0 \geq \theta_{\tau}(t_0) \ | \ T_i \geq t_0) = 1 - \tau$. As the underlying model for $\theta_{\tau}(t_0)$, we consider the following regression model with an exponential link: 
\begin{eqnarray} 
    \theta_{\tau} &=& \exp\left\{X_{i}^{\top}\beta_0(\tau, t_0)\right\}, \mbox{ or equivalently,} \nonumber \\
    \log(T_i - t_0) &=& X_{i}^{\top}\beta_0(\tau, t_0) + \epsilon_i \label{qr:mod2}
\end{eqnarray}
where $\beta_0(\tau, t_0)$ is a $(p+1) \times 1$ vector of regression coefficients and $\epsilon_i$ is a random variable taking zero at the $\tau$-th quantile. Note that when $t_0 = 0$, \eqref{qr:mod2} reduces to the quantile regression model for continuous responses \citep{koenker1978regression}. Hereafter, we use $\beta$ and $\theta$ instead of $\beta(\tau, t)$ and $\theta_{\tau}(t)$ for notational convenience. 

\subsection{Estimation via non-smooth functions} \label{sec:nonsmooth}

In the absence of censoring, the regression coefficients $\beta_0$ in \eqref{qr:mod2} could be estimated by solving the following estimating equations \citep{kim2012censored}
\begin{equation} \label{eq:ns:obj2}
    n^{-1}\sum_{i=0}^{n}I[T_i \ge t_0] X_i \Big\{I\big[\log(T_i - t_0) \leq X_i^{\top}\beta \big] - \tau \Big\} = 0
\end{equation}
Note that when $t_0 = 0$, Eq~\eqref{eq:ns:obj2} reduces to those developed for estimating the regression parameters in the quantile regression model for continuous responses \citep{koenker2005}.
    
In the presence of right-censorship, not all $T_i$s are observable. Thus, Eq~\eqref{eq:ns:obj2} cannot directly be evaluated. To account for this incompleteness in right-censored observations, weighted estimating equations in which a complete observation is weighted by the IPCW have been proposed \citep{li2016quantile}. The corresponding weighted estimating equations are 

\begin{equation} \label{eq:nsm:ipw}
    U_{t_0}(\beta, \tau) = n^{-1}\sum_{i=1}^{n}I[Z_i \ge t_0] X_i  \left\{I \big[\log(Z_i - t_0) \leq X_i^{\top}\beta \big]\frac{\delta_i}{\hat{G}(Z_i)/\hat{G}(t_0)}  -\tau \right\}
\end{equation}
where $\hat{G}(\cdot)$ is the Kaplan-Meier estimate of the survival function of the censoring time $C$. The estimator for $\beta_0$ in \eqref{qr:mod2}, $\hat{\beta}_{NS}$ is defined as the solution to \eqref{eq:nsm:ipw}. 

Note that \eqref{eq:nsm:ipw} are non-smooth step functions in $\beta$, whose exact solutions might not exist. Thus, instead of directly solving \eqref{eq:nsm:ipw}, $\hat{\beta}_{NS}$ can equivalently be obtained via minimizing $L_{t_0}(\beta, \tau)$, a $L_1$-objective function with the following form \citep{li2016quantile}: 
\begin{align} \label{l1:nsm}
    L_{t_0}(\beta, \tau) &= n^{-1}\sum_{i=1}^n \frac{\delta_i I[Z_i > t_0]}{\hat{G}(Z_i)/\hat{G}(t_0)} \left| \log(Z_i - t_0) - X_i^{\top}\beta \right| + \left| M - \beta^{\top}\sum_{l=1}^n -X_l \frac{\delta_l I[Z_l > t_0]}{\hat{G}(Z_l)/\hat{G}(t_0)}\right| \nonumber \\
    +& \ \left| M - \beta^{\top}\sum_{l=1}^n 2\tau X_l I[Z_l > t_0]\right|.
\end{align}
where $M$ is an extremely large positive constant (for example, $M = 10^6$) that bound $\left| \beta^{\top}\sum_{i=1}^n -X_i \frac{\delta_i I[Z_i > t_0]}{\hat{G}(Z_i)/ \hat{G}(t_0)}\right|$ and $\left| \beta^{\top}\sum_{i=1}^n 2\tau X_i I[Z_i > t_0]\right|$ from above. A straightforward calculation shows that the first derivative $L_{t_0}(\beta, \tau)$ with respect to $\beta$ is proportional to $U_{t_0}(\beta, \tau)$. $\hat{\beta}_{NS}$ can be easily obtained using some existing software that can implement $L_1$-minimization algorithm such as the \texttt{rq()} function in the \texttt{quantreg} package in R \citep{koenker2012package}.
    
\subsection{Estimation via induced smoothed functions} \label{sec:IS}

The estimating functions \eqref{eq:nsm:ipw} are non-smooth in regression coefficients. In this subsection, we propose to apply the induced smoothing procedure \citep{brown2005standard} to \eqref{eq:nsm:ipw}. The proposed induced smoothed estimating functions are constructed by adding a scaled mean-zero random noise to the regression parameters in \eqref{eq:nsm:ipw} and averaging it out. Specifically, 
\begin{align}\label{eq:is}
    \tilde{U}_{t_0}(\beta, \tau, H) & = E_w \{U_{t_0}(\beta+H^{1/2}W, \tau)\}\nonumber\\
    & = n^{-1} \sum_{i=1}^{n} I[Z_i \geq t_0] X_i \left\{ \Phi\Big(\frac{X_i^\top\beta-\log(Z_i-t_0)}{\sqrt{X_i^{\top} H X_{i}}}\Big)\frac{\delta_i}{\hat{G}(Z_i)/\hat{G}(t_0) }  -\tau \right\}
\end{align}
where $H = O(n^{-1})$, $W \sim N(0, I_p)$, $I_p$ is the $p \times p $ identity matrix, and $\Phi(\cdot)$ is the standard normal cumulative distribution function. The estimator for $\beta_0$ in \eqref{qr:mod2}, $\hat{\beta}_{IS}$, is defined as the solution to $\tilde{U}_{t_0}(\beta, \tau, H) = 0$. 

Note that \eqref{eq:is} is smooth in $\beta$, so calculation of $\hat{\beta}_{IS}$ can be readily done via the standard numerical algorithms such as the Newton-Raphson method. Moreover, since \eqref{eq:is} is differentiable with respect to $\beta$, the robust sandwich-type estimator, a typical approach employed in estimating equation-based approaches for variance estimation, can be directly applied; a slope matrix can be directly estimated. 
The resulting estimator is consistent and asymptotically normally distributed. In addition, as with other induced smoothed estimators for semiparametric AFT models \citep{johnson2009induced,pang2012variance,chiou2014fast,chiou2015semiparametric} and quantile regression models \citep{choi2018smoothed}, $n^{1/2}(\hat{\beta}_{IS} - \beta_0)$ and $n^{1/2}(\hat{\beta}_{NS} - \beta_0)$ are shown to be asymptotically equivalent, another important and useful feature of the induced smoothing method. These asymptotic properties are provided in Section~\ref{sec:asymp}.

\subsection{Variance estimation} \label{sec:var}

To estimate the variance-covariance function of $\hat{\beta}_{IS}$, we use the robust sandwich-form estimator, i.e., $\hat{\Var}(\hat{\beta}_{IS}, \tau) = \hat{A}(\hat{\beta}_{IS})^{\top} \hat{V}(\hat{\beta}_{IS}) \hat{A}(\hat{\beta}_{IS})$. For obtaining the slope matrix part, $\hat{A}(\beta)$, we take advantage of the smoothness of $\tilde{U}_{t_0}(\beta, \tau, H)$ in $\beta$ - $\hat{A}(\hat{\beta}_{IS})$ is the first derivative of $\tilde{U}_{t_0}(\beta, \tau, H)$ with respect to $\beta$ evaluated at $\hat{\beta}_{IS}$. Specifically,
\begin{align} \label{eq:cov:slp}
    \hat{A}(\hat{\beta}_{IS}) & = \frac{\partial \tilde{U}_{t_0}(\hat{\beta}_{IS}, \tau, H)}{\partial \beta} \nonumber \\
    & = n^{-1}\sum_{i=1}^{n} I[Z_i > t_0] X_i \frac{\hat{G}(t_0) \delta_i}{\hat{G}(Z_i)} \phi\bigg(\frac{{X_i}^{\top}\hat{\beta}_{IS} - \log(Z_i-t_0)}{\sqrt{{X_i}^{\top}H X_i}}\bigg)\bigg(\frac{-{X_i}}{\sqrt{{X_i}^{\top} H {X_i}}}\bigg)
\end{align}
where $\phi ( \cdot )$ is the density function of a standard normal distribution.
    
For calculating $\hat{V}(\beta)$, we propose to employ a computationally efficient resampling method. Similar procedures have been proposed for fitting semiparametric AFT models using the induced smoothing methods \citep{chiou2014fast}. In this procedure, we first generate $n$ independently and identically distributed ($i.i.d.$) positive multiplier random variables $\eta_i$'s, $i=1,...,n$, with both mean and variance being 1, independent of the observed data. Using generated multiplier, we update $\hat{G}^{\ast}(\cdot)$, a perturbed version of $\hat{G}(\cdot)$. Given data with a realization of $(\eta_1, \ldots, \eta_n)$, we obtain $\tilde{U}^{\ast}_{t_0}(\hat{\beta}_{IS}, \tau, H)$, a perturbed version of $\tilde{U}_{t_0}(\beta, \tau, H)$ where
\begin{equation} \label{eq:7}
    \tilde{U}^{\ast}_{t_0}(\beta, \tau, H) = n^{-1} \sum_{i=1}^{n} I[Z_i \geq t_0] X_i \eta_i \left\{ \Phi\left(\frac{X_i^\top\beta - \log(Z_i-t_0)}{\sqrt{X_i^{\top} H X_{i}}}\right)\frac{\hat{G}^{\ast}(t_0) \delta_i}{\hat{G}^{\ast}(Z_i)} -\tau \right\}
\end{equation}
We repeat this procedure $m$ times. Let $\tilde{U}^{\ast (k)}_{t_0}(\hat{\beta}_{IS}, \tau, H)$ denote the $k^{th}$ perturbed version of $\tilde{U}_{t_0}(\beta, \tau, H) (l=1, \ldots, m)$. Then, for given data, $\{\tilde{U}^{\ast (1)}_{t_0}(\hat{\beta}_{IS}, \tau, H), \ldots, \tilde{U}^{\ast (m)}_{t_0}(\hat{\beta}_{IS}, \tau, H)\}$ can be generated. 
$\hat{V}(\hat{\beta}_{IS})$ can be obtained by the sample variance of $\tilde{U}^{\ast}_{t_0}(\hat{\beta}_{IS}, \tau, H)$.

\section{Computation} \label{sec:comp}

For calculating $\hat{\beta}_{IS}$ and its estimated variance $\hat{\Var}(\hat{\beta}_{IS})$, we propose an iterative algorithm similar to those considered previously for the induced smoothing approach \citep{johnson2009induced, chiou2014fast,chiou2015semiparametric, choi2018smoothed}. The iterative algorithm uses the Newton-Raphson method while sequentially updating $\hat{\beta}_{IS}$ and $\hat{\Var}(\hat{\beta}_{IS})$ until convergence. This algorithm can be summarized as follows:

\begin{enumerate}
    \item[] {\bf Step 1}:
    Set the initial value as the non-smooth estimator for $\beta_0$ that minimizes \eqref{l1:nsm},  $\hat{\beta}^{(0)}=\hat{\beta}_{NS}, \hat{\Sigma}^{(0)} = I_{p}$ and $H^{(0)} = n^{-1}\hat{\Sigma}^{(0)}$.
	
	\item[] {\bf Step 2}: Given $\hat{\beta}^{(k)}$ and $H^{(k)}$ at the $k$-th step, update $\hat{\beta}^{(k)}$ by
	\begin{equation*}
	    \hat{\beta}^{(k+1)}=\hat{\beta}^{(k)} - \hat{A}(\hat{\beta}^{(k)})^{-1}{\tilde{U}_{t_0}(\hat{\beta}^{(k)}, \tau, H^{(k)}})
    \end{equation*}
	
	\item[] {\bf Step 3}: Given $\hat{\beta}^{(k+1)}$ and $\hat{\Sigma}^{(k)}$, update $\hat{\Sigma}^{(k)}$ by
	\begin{equation*}
	    \hat{\Sigma}^{(k+1)} = \hat{A}(\hat{\beta}^{(k+1)})^{-1} \hat{V}(\hat{\beta}^{(k+1)}, \tau) \hat{A}(\hat{\beta}^{(k+1)})^{-1}
	\end{equation*}
	
	\item[] {\bf Step 4}: Set $H^{(k+1)} = n^{-1}\hat{\Sigma}^{(k+1)}$. Repeat Steps 2 and 3 until $\hat{\beta}^{(k)}$ and $\hat{\Sigma}^{(k)}$ converge.
\end{enumerate}
$\hat{\beta}_{IS}$ and $\hat{\Sigma}$ are the values of $\hat{\beta}^{(k)}$ and $\hat{\Sigma}^{(k)}$ at convergence. $\hat{\Var}(\hat{\beta}_{IS}) = n^{-1}\hat{\Sigma}$.
    
In practice, a simpler version can be considered. Instead of updating $\hat{\Sigma}^{(k)}$ when calculating $\hat{\beta}^{(k)}$, we calculate $\hat{\beta}_{IS}$ at a fixed $H$, say $H = n^{-1}I_p$. Evaluating at $\hat{\beta}_{IS}$, $\hat{\Sigma}$ can be calculated using the variance estimation procedure described in Section 2.4. As shown in Section~\ref{sec:proof}, as long as $H = O(n^{-1})$, the choice of $H$ does not change the asymptotic properties of the resulting estimator. In addition, for induced smoothed estimators under semiparametric AFT models, the iterative algorithm and its simpler version were reported to produce similar estimates \citep{chiou2014fast,chiou2015semiparametric}. Our findings in the simulation studies in Section~\ref{sec:sim}, were also similar, so only the results from the simpler version are reported. 
	
\section{Asymptotic properties} \label{sec:asymp}

This section discusses the proposed estimator's asymptotic properties. We make several assumptions about regularity similar to those made in \citet{li2016quantile} and \citet{pang2012variance}. These conditions are required to establish the asymptotic properties of the proposed estimator. Under Conditions C1 - C3, it is possible to demonstrate that the proposed estimator is consistent and asymptotically normally distributed:
\begin{enumerate}
    \item[C1]
    \begin{flushleft}
        For any $t_0 \in \mathcal{T}$, the conditional density of $T-t_0$ given $T \geq t_0$, $g_{T-t_0}(s)$ is uniformly bounded from above and away from 0, and $g^{\prime}_{T-t_0}(s)$ exists and is uniformly bounded on the real line.\\
    \end{flushleft}
    \item[C2] For each $i=1,\dots, n, X_{i}$ satisfies the following conditions:
    \begin{flushleft}
        (a) $n^{-1}\Sigma_{i=1}^{n} X_{i} X_{i}^{\top}g_{T_i-t_0}(0)$ converges to a positive definite matrix A;\\
	    (b) There is a finite positive constant $M_c$ such that $\sup_{i}\lVert X_{i} \rVert \leq M_c$, where $\lVert \cdot \rVert$ denotes the Euclidean norm.\\
	\end{flushleft}
	\item[C3] There exists positive constant $\nu >0$ such that
    \begin{flushleft}
        (a) $P(C>\nu)=0$ and $P(C=\nu) \geq c_0$, where $c_0$ is some positive constant and\\
	    (b) $\sup_{X, t\in \mathcal{T}} [t + \exp(X^\top \beta)] \leq \nu$.\\
	\end{flushleft}
\end{enumerate}

The following theorem summarizes the asymptotic properties of the proposed estimator. 
\begin{theorem} \label{thm:1}
    Under Conditions C1 - C3, $\hat{\beta}_{IS}$ solving \eqref{eq:is} is consistent for $\beta_0$ and $n^{1/2}(\hat{\beta}_{IS}-\beta_0)$ is asymptotically normally distributed with mean zero and a finite covariance function matrix. Moreover, $n^{1/2}(\hat{\beta}_{IS}-\beta_0)$ has the same asymptotic distribution as that of $n^{1/2}(\hat{\beta}_{NS} -\beta_0)$ where $\hat{\beta}_{NS}$ minimizes \eqref{l1:nsm}.
\end{theorem}
A sketch of the proofs is provided in Appendix A.

\section{Simulation} \label{sec:sim}

We conduct comprehensive simulation tests to assess the suggested estimators' performance in finite samples. We use simulation settings similar to those in \citet{jung2009regression}. We assume the proposed model (\ref{qr:mod2})  includes a single binary covariate with success probability $0.5$. We generate $T$ from a Weibull distribution with the survival function $S(t) = \exp\{-(\rho t)^\kappa\}$. We set the scale parameter $\kappa$ to $2$. For $t_0 = 0$, we set the intercept $\beta^{(0)}(\tau, t_0) = \log(5)$. We consider two settings for the regression coefficient for $X$, $\beta^{(1)}$: $\beta^{(1)}(\tau, t_0) = 0$ and $\beta^{(1)}(\tau, t_0) \neq 0$. For $\beta^{(1)}(\tau, t_0) \neq 0$ with $t_0 = 0$, we set $\beta^{(1)}(\tau, t_0) = \log(2)$. For a given $\tau$, $\beta^{(0)}(\tau, t_0)$ and $\beta^{(1)}(\tau, t_0)$ at $t_0 = 0$, the shape parameter $\rho$ can be obtained by solving 
\begin{equation} \label{eq:sim:weibull}
    \rho^{-1}\{ (\rho t_0)^\kappa - \log (1-\tau) \}^{(1/\kappa)}- t_0 = \exp\{\beta^{(0)}(\tau, t_0) + \beta^{(1)}(\tau, t_0)\}.    
\end{equation}
Let $\rho_0(\tau)$ and $\rho_1(\tau)$ denote $\rho$s at a given $\tau$ for $\beta^{(1)}(\tau, t_0) = 0$ and $\beta^{(1)}(\tau, t_0) \neq 0$, respectively. When $t_0 > 0$, for a given $\tau$ and $\kappa$, $\beta^{(0)}(\tau, t_0)$s and $\beta^{(1)}(\tau, t_0)$s can be obtained by solving \eqref{eq:sim:weibull} sequentially for $\beta^{(0)}(\tau, t_0)$ by setting $\beta^{(1)}(\tau, t_0) = 0$ and then for $\beta^{(1)}(\tau, t_0)$ at the given $\beta^{(0)}(\tau, t_0)$. We consider $t_0 = 0, 1, 2$ and $3$ for $t_0$ and $\tau = 0.25$ and $0.5$ for $\tau$. For $\tau = 0.5$, the corresponding $\beta^{(0)}(\tau, t_0)$s and $\beta^{(1)}(\tau, t_0)$s are $1.61(= \log(5)), 1.41, 1.22, 1.04$ and $0.69(=\log(2)), 0.80, 0.91, 1.02$, respectively.
    
Potential censoring times, $C_i$s, are generated, independently from $T_i$s, from $unif(0, c)$ where $c$ is determined by desired censoring proportions. We consider censoring proportions of $0\%, 10\%, 30\%$ and $50\%$. The sample size is set to $n=200$. For variance estimation, the resampling size for estimating $\hat{V}$ is set to $200$. The estimates obtained for each configuration are based on $2000$ repetitions. 
    
Simulation results for $\tau = 0.5$ when $\beta^{(1)} = 0$ are summarized in Table~\ref{table:sim0}.
    
\begin{table}[htp]
\caption{Simulation results of fitting quantile regression model for residual lifetimes using the proposed induced smoothing method at $\tau = 0.5$. $\beta^{(0)}(0.5,t_0) = 1.61, 1.41, 1.22, 1.04$ at $t_0 = 0, 1, 2, 3$ and $\beta^{(1)}(0.5,t_0) = 0$. PE is the mean of point estimates for regression parameters, $\beta^{(0)}(\tau,t_0)$ and $\beta^{(1)}(\tau,t_0)$. ESE is the mean of estimated standard error of regression parameter. SD is the sample standard deviation of point estimates. The coverage proportion of the nominal 95\% confidence intervals is denoted by CP. Cens is a term that refers to the proportions of censorship. The sample size has been determined to be 200. The number of repetition is 2000.}
\centering
\begin{tabular}{|c|c|c|c|c|c|c|c|c|c|}
    \hline
    \multirow{2}{*}{$t_0$} & \multirow{2}{*}{Cens} & \multicolumn{4}{c|}{$\beta^{(0)}(0.5, t_0)$} & \multicolumn{4}{c|}{$\beta^{(1)}(0.5, t_0)$}\\ \cline{3-10}
	& & PE & ESE & SD & CP & PE & ESE & SD & CP\\
	\hline\hline
	\multirow{4}{*}{0} & 0 & 1.608 & 0.068 & 0.069 & 0.927 & -0.003 & 0.097 & 0.095 & 0.949 \\ 
    & 10 & 1.608 & 0.072 & 0.073 & 0.931 & -0.002 & 0.104 & 0.101 & 0.940 \\ 
    & 30 & 1.609 & 0.079 & 0.081 & 0.926 & -0.002 & 0.118 & 0.116 & 0.934 \\ 
    & 50 & 1.610 & 0.091 & 0.091 & 0.915 & -0.003 & 0.138 & 0.135 & 0.935 \\
	\hline
	\multirow{4}{*}{1} & 0 & 1.408 & 0.084 & 0.084 & 0.926 & -0.003 & 0.120 & 0.115 & 0.947 \\ 
    & 10 & 1.408 & 0.088 & 0.088 & 0.926 & -0.002 & 0.128 & 0.123 & 0.940 \\ 
    & 30 & 1.410 & 0.099 & 0.100 & 0.919 & -0.003 & 0.146 & 0.143 & 0.934 \\ 
    & 50 & 1.411 & 0.115 & 0.117 & 0.907 & -0.002 & 0.175 & 0.170 & 0.931 \\ 
	\hline
	\multirow{4}{*}{2} & 0 & 1.215 & 0.100 & 0.101 & 0.914 & 0.000 & 0.144 & 0.139 & 0.941 \\ 
    & 10 & 1.214 & 0.108 & 0.107 & 0.913 & 0.002 & 0.156 & 0.150 & 0.941 \\ 
    3& 30 & 1.216 & 0.122 & 0.124 & 0.905 & 0.002 & 0.181 & 0.179 & 0.932 \\ 
    & 50 & 1.222 & 0.144 & 0.150 & 0.879 & -0.002 & 0.219 & 0.216 & 0.917 \\  
	\hline
	\multirow{4}{*}{3} & 0 & 1.035 & 0.121 & 0.120 & 0.916 & 0.001 & 0.172 & 0.168 & 0.938 \\ 
    & 10 & 1.034 & 0.131 & 0.131 & 0.904 & 0.003 & 0.190 & 0.185 & 0.932 \\ 
    & 30 & 1.036 & 0.153 & 0.153 & 0.885 & 0.003 & 0.228 & 0.221 & 0.920 \\ 
    & 50 & 1.037 & 0.183 & 0.191 & 0.854 & 0.010 & 0.287 & 0.274 & 0.898 \\ 
	\hline
\end{tabular}
\label{table:sim0}
\end{table}

Overall, our proposed estimators work reasonably well in most cases considered. The point estimates are all close to the true regression parameters and their standard errors estimates are virtually identical to their empirical counterparts. The coverage rates of the nominal $95\%$ confidence intervals are in the range of $93\%$ to $95\%$ except for $t_0 = 2$ and $3$, and especially when the censoring proportion is $50\%$. This is mainly due to decreased effective sample sizes. When $t_0 = 2$ and $3$, the number of effective sample sizes decrease to, on average, $168$ and $146$, respectively. When we increase the sample size to $400$, the coverage rates get closer to the nominal level of $95\%$ (Table 3 in Supplementary material). Calculating point and standard error estimates for a single data set takes an average of $0.218$ seconds using the proposed induced smoothing method. It is approximately $14\%$ and $77\%$ faster than the non-smooth estimator and the iterative smoothed estimator, respectively. All analyses were run on a 4.2 GHz Intel(R) quad Core(TM) i7-7700K central process unit (CPU) and conducted by R 4.02 \citep{r2020}. The proposed methods are implemented in a \textbf{R} package \texttt{qris}, which is freely available at \emph{https://github.com/Kyuhyun07/SQRL}. 

The simulation results for $\tau = 0.5$ when $\beta^{(1)} \neq 0$ are presented in Table~\ref{table:simnot_0}. The overall findings are similar to those under $\beta^{(1)} = 0$. The coverage rates of the nominal $95\%$ confidence intervals are relatively low when $t_0 = 2$ and $3$ with the 50\% censoring proportion. Again, by increasing the sample size to $400$, the coverage rates become closer to the $95\%$ level in Table 6 in Supplementary material.

\begin{table}[htp]
\caption{Simulation results of fitting quantile regression model for residual lifetimes using the proposed induced smoothing method at $\tau = 0.5$. $\beta^{(0)}(0.5,t_0) = 1.61, 1.41, 1.22, 1.04$ and $\beta^{(1)}(0.5,t_0) = 0.69, 0.80, 0.91, 1.02$ at $t_0 = 0, 1, 2, 3$. PE is the mean of point estimates for regression parameters, $\beta^{(0)}(\tau,t_0)$ and $\beta^{(1)}(\tau,t_0)$. ESE is the mean of estimated standard error of regression parameter. SD is the sample standard deviation of point estimates. The coverage proportion of the nominal 95\% confidence intervals is denoted by CP. Cens is a term that refers to the proportions of censorship. The sample size has been determined to be 200. The number of repetition is 2000.}
\centering
\begin{tabular}{|c|c|c|c|c|c|c|c|c|c|}
    \hline
    \multirow{2}{*}{$t_0$} & \multirow{2}{*}{Cens} & \multicolumn{4}{c|}{$\beta^{(0)}(0.5, t_0)$} & \multicolumn{4}{c|}{$\beta^{(1)}(0.5, t_0)$}\\ \cline{3-10}
    & & PE & ESE & SD & CP & PE & ESE & SD & CP\\
    \hline\hline
    \multirow{4}{*}{0} & 0 & 1.608 & 0.068 & 0.069 & 0.928 & 0.690 & 0.097 & 0.095 & 0.948 \\ 
    & 10 & 1.608 & 0.070 & 0.072 & 0.921 & 0.691 & 0.103 & 0.101 & 0.941 \\ 
    & 30 & 1.607 & 0.075 & 0.076 & 0.931 & 0.693 & 0.117 & 0.116 & 0.942 \\ 
    & 50 & 1.609 & 0.082 & 0.083 & 0.927 & 0.694 & 0.138 & 0.135 & 0.928 \\ 
	\hline
	\multirow{4}{*}{1} & 0 & 1.408 & 0.083 & 0.084 & 0.927 & 0.789 & 0.114 & 0.110 & 0.952 \\ 
    & 10 & 1.408 & 0.086 & 0.087 & 0.926 & 0.789 & 0.121 & 0.118 & 0.939 \\ 
    & 30 & 1.408 & 0.093 & 0.093 & 0.927 & 0.792 & 0.137 & 0.135 & 0.935 \\ 
    & 50 & 1.409 & 0.100 & 0.101 & 0.917 & 0.792 & 0.157 & 0.158 & 0.908 \\ 
	\hline
	\multirow{4}{*}{2} & 0 & 1.215 & 0.100 & 0.101 & 0.913 & 0.883 & 0.133 & 0.127 & 0.941 \\ 
    & 10 & 1.215 & 0.104 & 0.106 & 0.911 & 0.884 & 0.141 & 0.137 & 0.932 \\ 
    & 30 & 1.215 & 0.113 & 0.115 & 0.910 & 0.886 & 0.160 & 0.158 & 0.928 \\ 
    & 50 & 1.216 & 0.126 & 0.126 & 0.902 & 0.882 & 0.184 & 0.188 & 0.899 \\ 
	\hline
	\multirow{4}{*}{3} & 0 & 1.035 & 0.121 & 0.120 & 0.918 & 0.966 & 0.155 & 0.147 & 0.929 \\ 
    & 10 & 1.035 & 0.125 & 0.126 & 0.903 & 0.967 & 0.164 & 0.159 & 0.925 \\ 
    & 30 & 1.035 & 0.138 & 0.139 & 0.901 & 0.968 & 0.189 & 0.184 & 0.917 \\ 
    & 50 & 1.035 & 0.152 & 0.151 & 0.881 & 0.964 & 0.213 & 0.210 & 0.898 \\ 
	\hline
\end{tabular}
\label{table:simnot_0}
\end{table}

We also consider a lower quantile with $\tau = 0.25$ when $\beta^{(1)} \neq 0$. Simulation results under $\tau=0.25$ are presented in Table~\ref{table:simnot_025}. 

\begin{table}[ht]
\caption{Simulation results of fitting quantile regression model for residual lifetimes using the proposed induced smoothing method at $\tau=0.25$. $\beta^{(0)}(0.25,t_0) = 1.61, 1.41, 1.22, 1.04$ and $\beta^{(1)}(0.25,t_0) = 0.69, 0.80, 0.91, 1.02$ at $t_0 = 0, 1, 2, 3$. PE is the mean of point estimates for regression parameters, $\beta^{(0)}(\tau,t_0)$ and $\beta^{(1)}(\tau,t_0)$. ESE is the mean of estimated standard error of regression parameter. SD is the sample standard deviation of point estimates. The coverage proportion of the nominal 95\% confidence intervals is denoted by CP. Cens is a term that refers to the proportions of censorship. The sample size has been determined to be 200. The number of repetition is $2000$.}
\centering
\begin{tabular}{|c|c|c|c|c|c|c|c|c|c|}
	\hline
	\multirow{2}{*}{$t_0$} & \multirow{2}{*}{Cens} & \multicolumn{4}{c|}{$\beta^{(0)}(0.25, t_0)$} & \multicolumn{4}{c|}{$\beta^{(1)}(0.25, t_0)$}\\ \cline{3-10}
	& & PE & ESE & SD & CP & PE & ESE & SD & CP\\
	\hline\hline
	\multirow{4}{*}{0} & 0 & 1.608 & 0.096 & 0.097 & 0.906 & 0.683 & 0.137 & 0.134 & 0.942 \\ 
    & 10 & 1.608 & 0.097 & 0.099 & 0.905 & 0.684 & 0.141 & 0.136 & 0.941 \\ 
    & 30 & 1.608 & 0.101 & 0.102 & 0.902 & 0.685 & 0.150 & 0.145 & 0.941 \\ 
    & 50 & 1.609 & 0.105 & 0.106 & 0.905 & 0.687 & 0.163 & 0.154 & 0.931 \\ 
	\hline
	\multirow{4}{*}{1} & 0 & 1.409 & 0.116 & 0.119 & 0.896 & 0.782 & 0.160 & 0.157 & 0.939 \\ 
    & 10 & 1.409 & 0.119 & 0.121 & 0.892 & 0.782 & 0.165 & 0.159 & 0.939 \\ 
    & 30 & 1.408 & 0.123 & 0.125 & 0.893 & 0.784 & 0.175 & 0.169 & 0.941 \\ 
    & 50 & 1.409 & 0.129 & 0.130 & 0.896 & 0.787 & 0.191 & 0.181 & 0.932 \\ 
	\hline
	\multirow{4}{*}{2} & 0 & 1.216 & 0.137 & 0.137 & 0.896 & 0.874 & 0.183 & 0.176 & 0.928 \\ 
    & 10 & 1.217 & 0.139 & 0.140 & 0.891 & 0.875 & 0.189 & 0.179 & 0.933 \\ 
    & 30 & 1.217 & 0.145 & 0.145 & 0.892 & 0.876 & 0.201 & 0.193 & 0.928 \\ 
    & 50 & 1.218 & 0.153 & 0.154 & 0.881 & 0.879 & 0.221 & 0.209 & 0.922 \\ 
	\hline
	\multirow{4}{*}{3} & 0 & 1.036 & 0.158 & 0.159 & 0.888 & 0.957 & 0.208 & 0.199 & 0.910 \\ 
    & 10 & 1.036 & 0.159 & 0.161 & 0.890 & 0.958 & 0.212 & 0.201 & 0.916 \\ 
    & 30 & 1.036 & 0.166 & 0.168 & 0.881 & 0.960 & 0.228 & 0.217 & 0.906 \\ 
    & 50 & 1.035 & 0.175 & 0.178 & 0.876 & 0.965 & 0.249 & 0.236 & 0.911 \\ 
	\hline
\end{tabular}
\label{table:simnot_025}
\end{table}

In general, the proposed estimator appears to perform well in this situation, with little bias in the point estimates and little discrepancies between the proposed standard errors estimates and their empirical counterparts. The coverage proportions are, however, low in the range of $88\%$ to $91\%$ for $\beta^{(0)}(0.25, t_0)$ when $t_0 \neq 0$. Again, raising the sample size to $400$ shows an improvement in the coverage rates, which approach $95\%$. These outcomes are summarized in Table 5 in Supplementary material. Additionally, simulation results for various quantiles with $\tau = 0.25$ and $\tau = 0.75$ with 200 sample size are included in Tables 1, 2, and 4 in Supplementary material. 

We also consider simulation settings under high censoring rate. Other settings remain identical to those used in earlier simulations. The results for the censoring rates of $70\%$ in different quantiles are summarized in Supplementary material Tables 8 and 9. For $\tau = 0.25$ and $t_0 = 0$, the proposed induced smoothed estimator appears to perform reasonably well. In general, however, the results demonstrate large biases and low coverage rates. Notably, the higher quantiles, $\tau = 0.75$, may not be identifiable for failure times with high censoring rates. Additional simulation results for the censoring rates of $60\%$ and $80\%$ with $\tau = 0.5$ are included in Supplementary material Table 10. As expected, the proposed estimator performs poorly except those with a 60\% censoring rate and $t_0 = 0$.

\begin{figure*}[htp]
\centering
\subfigure[$\beta^{(0)}(0.5, 0) = 1.61$]{
    \includegraphics[width=.45\columnwidth]{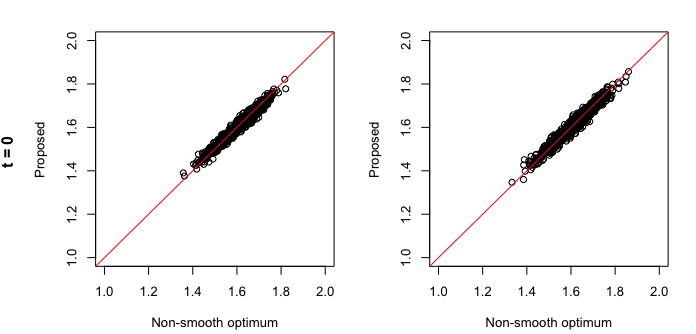}
    \label{fig:beta_0_0}
    }
\subfigure[$\beta^{(1)}(0.5, 0) = 0.69$]{
    \includegraphics[width=.45\columnwidth]{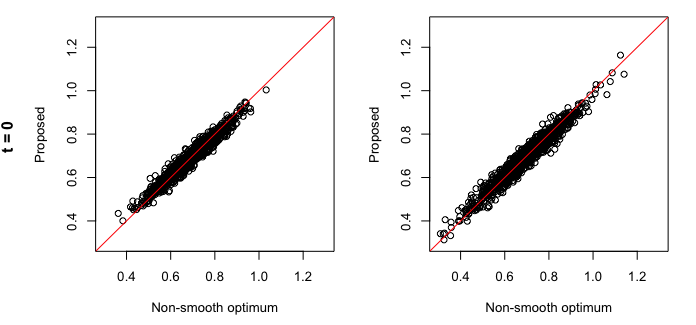}
    \label{fig:beta_1_0}
    }	
\subfigure[$\beta^{(0)}(0.5, 2) = 1.215$]{
    \includegraphics[width=.45\columnwidth]{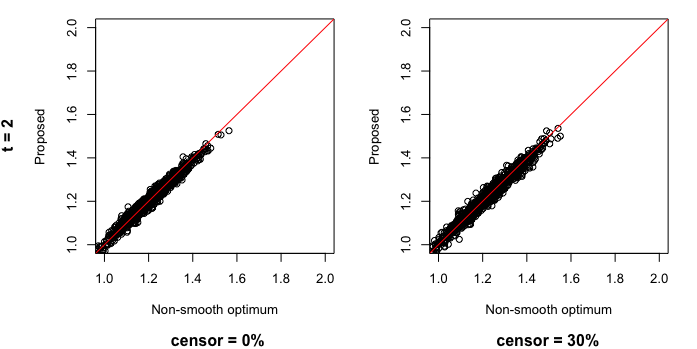}
    \label{fig:beta_0_2}
    }
\subfigure[$\beta^{(1)}(0.5, 2) = 0.883$]{
    \includegraphics[width=.45\columnwidth]{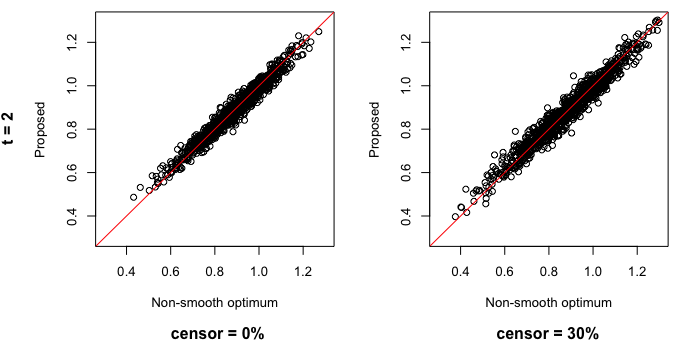}
    \label{fig:beta_1_2}
    }	
\caption{Scatter plots of regression coefficients estimates. The horizontal and vertical axes represent the estimates from the non-smooth and proposed methods, respectively. The circle represents a pair of the two point estimates ($\hat{\beta}_{NS}$, $\hat{\beta}_{IS}$) and The solid line represents the line with the 45 degree angle going through the origin.}
\label{fig:scattered}
\end{figure*}

We also compare the proposed induced smoothed estimator with its non-smooth counterpart \citep{li2016quantile}. Under the setting previously considered for $\tau = 0.5$ with nonzero $\beta^{(1)}(\tau, t_0)$, we calculated the non-smooth estimates. The non-smoothed estimator is obtained via the $L_1$-minimization method in \eqref{l1:nsm} using the \texttt{rq()} function in the \texttt{quantreg} package \citep{koenker2012package}. To assess the degree to which the two methods coincide with each other, we draw scatter plots comparing the proposed induced-smoothed estimates to the non-smoothed estimates for $\beta^{(0)}$ and $\beta^{(1)}$s at each combination of different $t_0$s ($t_0=0$ and $2$) and censoring proportions ($0\%$ and $30\%$). Scatter plots comparing the suggested induced smoothed and non-smooth estimates for various combinations are shown in Figure~\ref{fig:scattered}.

Each plot demonstrates that pairs of the two estimates (circles) are distributed about the straight line with a 45 degree angle (red line), indicating that the two methods give similar estimates in general.  The empirical standard errors for $2000$ estimates are similar, but those from the proposed estimator are consistently smaller by 3\% to 5\% (Table 7,  Supplementary material). Similar conclusions can be seen in a paper that compares non-smooth and induced smoothed estimators for quantile regression models with $t_0 = 0$, a particular example of the scenario we consider, but for competing risks data \citep{choi2018smoothed}. 

We conclude this section with some remarks. First, presented simulation results are based on the non-iterative algorithm using $H = I_2$. While the iterative method produces slightly superior results in some cases, the non-iterative and iterative algorithms perform similarly in the majority of the scenarios we investigated.
Second, the covariate considered in the simulation experiments is binary. We also consider a continuous covariate following a $unif(0, 1)$ distribution. Overall, the performances are comparable to those obtained using a binary covariate. The setting and simulation results are described in Supplementary material (Tables 11 and 12). 

\section{Real data analysis} \label{sec:realdata}

We illustrate our proposed methods by analyzing dental restoration longevity study data from a cohort of older adults \citep{caplan2019dental}. Dental caries is known to be the most frequently encountered health condition worldwide, including among older adult populations \citep{kassebaum2015global}. As the older adult population grows, a larger burden and impact on society and the health care system are expected. A recent study addresses the importance of dental restoration longevity among older adults and evaluated an extended Cox regression model for longevity of dental restorations and related factors \citep{caplan2019dental}. The data set arose from patients $\geq$65 years of age treated at the University of Iowa College of Dentistry who had dental restorations of different types and sizes placed during the years 2000-2014 (where ``dental restoration" is a general term that refers to repairing a damaged tooth).  For a more detailed description of the data set, see \citet{caplan2019dental}. Patients were followed until they incurred an event, which in this study was represented by restoration replacement, extraction of the tooth, or endodontic treatment of the tooth.  We selected a sample from these patients, and because a patient could have had multiple teeth restored, we randomly choose the first restored tooth for each patient. The resulting sample had $1551$ unique patients who contributed one restored tooth. The failure time for a tooth was defined as the longevity of the initial restoration of the tooth, i.e., time from the initial restoration to the subsequent restoration or extraction of the tooth. If a tooth had not received any subsequent restoration nor had been extracted at the last visit, failure time was considered as censored at the last visit. The corresponding censoring proportion was $56.6\%$. For factors that might be related to the longevity of the initial restoration of the tooth, we considered the following 7 variables as covariates: gender(male/female), age at baseline, cohort effects (4, 5 and 6 for the group of patients who received treatment between 2000 and 2004, between 2005 and 2009, and between 2010 and 2014, respectively), provider type (predoctoral student / graduate student or faculty), payment (private / Medicaid (Title XIX) / self-pay), tooth type (molar/premolar/anterior), and restoration type (amalgam / composite / GIC / crown or bridge).

We use a semiparametric quantile regression model to examine the effect of the aforementioned variables on the quantiles of longevity following dental restoration at various time points including the baseline ($t_0 = 0$) and subsequent follow-up visits ($t_0 = 1$ and $2$ years). As a reference group, we included female patients who underwent their initial restoration between $2010$ and $2014$ (cohort$=6$) in an anterior tooth with crown and bridges by a graduate student or faculty member and were self-paying.

Table~\ref{table:realdata} summarizes the model fitting findings at the $\tau = 0.1$ and $0.2$ quantiles and various follow-up times ($t_0 = 0, 1$, and 2 years). Note that we focus on lower quantiles that can be confidently calculated due to the identifiability issue caused by the high censoring rate. Regression coefficients are estimated by the proposed induced smoothing approach. The suggested robust sandwich-type estimator is used to estimate the estimators' standard errors, supported by a resampling-based technique.

\begin{table}[ht]
\caption{Results of analyzing the dental restoration data for $\tau=0.1$ and $0.2$ quantiles of residual longevity after dental restoration at $t_0 = 0, 1$, and $2$ (years), respectively. PE is point estimate of the regression parameter. SE is the estimated standard error of the regression parameter estimator.} 
\small
\centering
\begin{tabular}{|c|c|c|c|c|c|c|c|c|c|c|c|c|}
    \hline
    \multirow{4}{*}{Covariate} & \multicolumn{12}{c|}{$\tau$}\\
    \cline{2-13}
    & \multicolumn{6}{c|}{$0.1$} & \multicolumn{6}{c|}{$0.2$} \\
    \cline{2-13}
    & \multicolumn{2}{c|}{$t_0=0$} & \multicolumn{2}{c|}{$t_0=1$} & \multicolumn{2}{c|}{$t_0=2$} & \multicolumn{2}{c|}{$t_0=0$} & \multicolumn{2}{c|}{$t_0=1$} & \multicolumn{2}{c|}{$t_0=2$} \\ 
    \cline{2-13}
    & PE & SE & PE & SE & PE & SE & PE & SE & PE & SE & PE & SE\\
    \hline
    Male & -0.076 & 0.177 & 0.175 & 0.261 & -0.704 & 0.339 & -0.011 & 0.147 & 0.163 & 0.236 & -0.463 & 0.324\\ 
    Age & 0.014 & 0.010 & -0.022 & 0.019 & -0.035 & 0.028 & 0.030 & 0.011 & -0.004 & 0.016 & -0.026 & 0.025\\ 
    Cohort4 & -0.577 & 0.241 & -0.163 & 0.276 & 0.503 & 0.433 & -0.793 & 0.226 & -0.142 & 0.274 & 0.336 & 0.824\\ 
    Cohort5 & -0.606 & 0.311 & -0.624 & 0.275 & 0.253 & 0.494 & -0.673 & 0.176 & -0.371 & 0.299 & 0.116 & 0.773\\ 
    Predoc & 0.170 & 0.230 & 0.482 & 0.227 & -0.042 & 0.384 & 0.381 & 0.133 & 0.036 & 0.395 & -0.145 & 0.590\\ 
    Private & 0.082 & 0.223 & 0.057 & 0.292 & -0.132 & 0.510 & 0.390 & 0.194 & -0.051 & 0.253 & -0.254 & 0.552\\ 
    XIX & -0.132 & 0.263 & -0.008 & 0.256 & -0.083 & 0.346 & 0.243 & 0.215 & -0.308 & 0.321 & -0.190 & 0.640\\ 
    Molar & 0.205 & 0.212 & 0.012 & 0.372 & 0.372 & 0.766 & 0.093 & 0.197 & 0.010 & 0.348 & -0.038 & 0.388\\ 
    Pre-molar & 0.082 & 0.173 & 0.910 & 0.208 & -0.324 & 0.395 & 0.126 & 0.180 & 0.669 & 0.244 & -0.214 & 0.454\\ 
    Amalgam & -2.101 & 1.317 & -1.975 & 0.406 & -0.198 & 1.027 & -2.525 & 0.291 & -1.656 & 0.641 & -0.373 & 0.544\\ 
    Composite & -2.312 & 1.288 & -2.160 & 0.406 & -0.055 & 1.140 & -2.626 & 0.286 & -1.743 & 0.713 & -0.470 & 0.770\\ 
    GIC & -2.001 & 1.318 & -2.167 & 0.450 & -1.015 & 1.065 & -2.303 & 0.323 & -1.948 & 0.650 & -0.858 & 0.549\\ 
    \hline
\end{tabular}
\label{table:realdata}
\end{table}
    
The results show that the effects of factors vary depending on the quantiles considered or the time for defining residual life. For example, when $\tau = 0.1$, i.e., the $10^{th}$ percentile of the residual longevity of the restored teeth, at the 5\% significance level, the residual longevity of the restored tooth at baseline or one year after the initial restoration for a male does not appear to be statistically significantly different from that of a female while the remaining covariates are held constant. However, when evaluating the residual longevity of the restored tooth at 2 years after the initial restoration, a male is estimated to be 0.7 years shorter in the logarithm scale than a female, a statistically significant effect ($p$-value $= 0.038$). A similar pattern can be observed when $\tau = 0.2$, but none of the effects appear to be statistically significant. Other results at different quantiles $\tau = 0.05, 0.15, 0.25$ are shown in Supplementary Tables 13, 14 and 15.

\begin{figure*}[ht] 
\centering
    \subfigure[]{
    \includegraphics[scale=0.45]{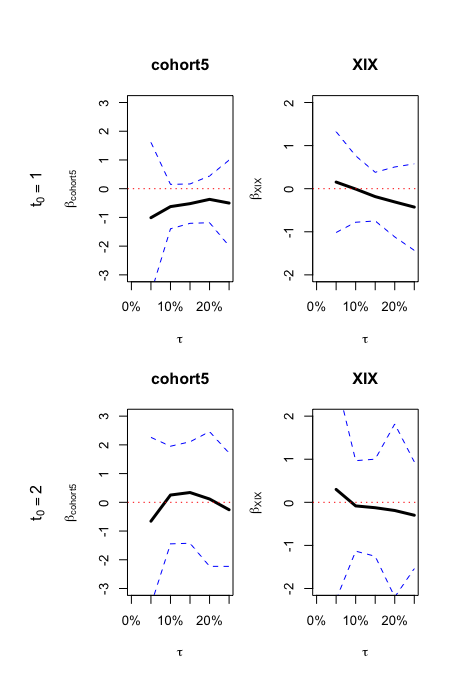}
    \label{fig:CI-quantile}
    }
    \subfigure[]{
	\includegraphics[scale=0.45]{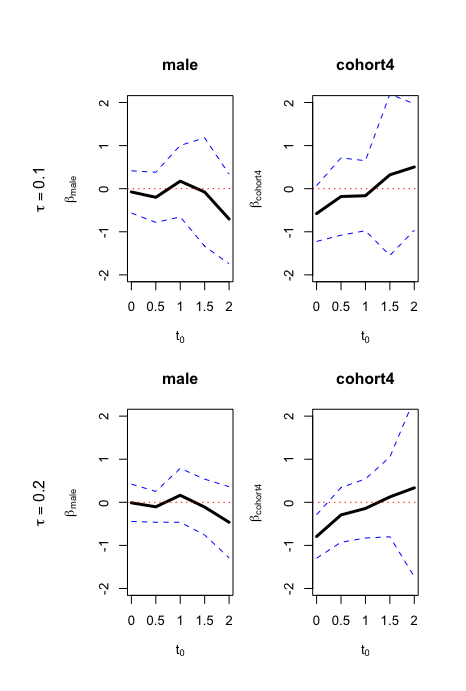}
	\label{fig:CI-t_0}
	}	
\caption{(a) Estimated effects of cohort5 (patients treated between 2005 and 2009) and XIX (Medicaid) along with corresponding 95\% pointwise confidence intervals for quantiles ranging from 0.05 to 0.25 at $t_0=1$ and 2. The black solid lines represent the point estimates of the regression parameters for cohorts 5 and XIX, denoted by the variables $\beta_{cohort5}$ and $\beta_{XIX}$. The blue dotted lines represent the upper and lower limits of the $95 percent $ pointwise confidence intervals for $\beta_{cohort5}$ and $\beta_{XIX}$, respectively. 
(b) Estimated effects of male and cohort4 (patients treated between 2000 and 2004) along with corresponding 95\% pointwise confidence intervals for times ranging from 0 to 2 at $\tau=0.1$ and $0.2$. The black solid lines represent the point estimates of the regression parameters for male and cohort4, $\beta_{male}$ and $\beta_{cohort4}$. The blue dotted lines represent the upper and lower limits of the $95\%$ pointwise confidence intervals for $\beta_{male}$ and $\beta_{cohort4}$}
\label{fig:CI}
\end{figure*}

To illustrate these variable effects, we plot the predicted regression coefficients for several covariates by different quantiles and times in Figure~\ref{fig:CI}, together with the accompanying 95\% Wald-type pointwise confidence intervals. For instance, at $t_0 = 1$, the estimated effects of Medicaid (XIX) on self-pay demonstrate a diminishing tendency as quantiles are raised. A similar pattern can be observed at $\tau = 0.2$, although with a slower falling tendency.

\section{Discussion} \label{sec:discussion}
	
In this paper, we offer a strategy for fitting a semiparametric quantile regression model for residual life subject to right-censoring. Existing estimation approaches \citep{jung2009regression,kim2012censored,li2016quantile} are based on estimating equations non-smooth in regression parameters. Our suggested technique for estimating regression parameters adapts the induced smoothing procedure, which has been demonstrated to be computationally efficient and reliable when used with semiparametric AFT models or quantile regression \citep{chiou2014fitting, chiou2015semiparametric,choi2018smoothed}. The standard error of the proposed estimator can be estimated by the robust sandwich-type estimator with the application of an efficient resampling method. The proposed estimator is shown to have desirable asymptotic properties: consistent and asymptotically normal. Extensive simulation experiments show that the proposed estimator performs reasonably well for finite samples.
	
\citet{kim2012censored} also considered a similar problems and proposed estimating equations with the following form: 
\begin{equation} \label{eq:nsm:kim}
    U_n(\beta) = n^{-1}\sum_{i=1}^{n}I[Z_i \ge t_0] X_i \frac{\delta_i}{\hat{G}(Z_i)} \left(I[\log(Z_i - t_0) \leq X_i^{\top}\beta]  -\tau \right)
\end{equation}
Except for the position of the IPCW, \eqref{eq:nsm:kim} is the same as \eqref{eq:nsm:ipw}, on which our proposed estimating equations are based. An alternative estimation procedure can also be considered by applying the induced smoothing approach to \eqref{eq:nsm:kim}. This alternative version can be directly applied to the proposed computing algorithm and standard error estimation procedure. The asymptotic properties can be obtained using arguments in \ref{sec:proof} with a minimal modification. Through simulation experiments, we also considered this alternative version and compared it to our proposed estimator. Overall, the results are comparable. However, as the censoring proportion increases, the alternative version's performance falls behind that of the proposed estimator. This phenomenon warrants further investigation.


\section*{Appendix A. Proof of Theorem 1} \label{sec:proof}

In this Appendix, we provide a proof of Theorem 1: consistency and asymptotic normality of the proposed induced smoothed estimator.

First, we establish the consistency of the proposed estimator $\hat{\beta}_{IS}$. The consistency of the non-smooth counterpart, $\hat{\beta}_{NS}$, is shown in \citep{li2016quantile}. Based on this consistency result, it suffices to we prove that, as $n \to \infty$, the difference between $\tilde{U}_{t_0}(\beta, \tau, H)$ and $U_{t_0}(\beta, \tau)$ scaled by $n^{1/2}$ converges uniformly to zero in probability for $\beta$ in the compact neighborhood of $\beta_0$. 

Let $\sigma_i=(X_i^\top H X_i)^{1/2}$, $\epsilon_i(\beta) = X_i\beta - \log(Z_i-t_0)$ and $d_i(\beta) = \sign (\epsilon_i^\beta)\Phi(-\lvert\epsilon_i^\beta/\sigma_i\rvert)$. 
Then,
\begin{align*} \label{eq:cons:diff1}
& n^{1/2}\{\tilde{U}_{t_0}(\beta, \tau, H) - U_{t_0}(\beta, \tau))\}\\
& = {n^{-1/2}} \sum_{i=1}^{n}I[Z_i \geq t_0] X_i \delta_i \frac{\hat{G}(t_0)}{\hat{G}(Z_i)} \bigg\{ \Phi \bigg(\frac{-\epsilon_i(\beta)}{\sigma_i}\bigg)-I[\epsilon_i(\beta) < 0] \bigg\}\\ 
& = n^{-1/2} \sum_{i=1}^{n}I[Z_i \geq t_0] X_i \delta_i \frac{G(t_0)}{G(Z_i)}d_i(\beta) + n^{-1/2} \sum_{i=1}^{n}I[Z_i \geq t_0] X_i \delta_i \left\{\frac{\hat{G}(t_0)}{\hat{G}(Z_i)} - \frac{G(t_0)}{G(Z_i)}\right\} d_i(\beta)\\
& = \ D^{(1)}_{n}(\beta) + D^{(2)}_{n}(\beta)
\end{align*}

To show $\lVert D^{(1)}_n(\beta) \rVert \xrightarrow{p} 0$ as $n \to \infty$, we first note that
\begin{align*}
\E\{D^{(1)}_n(\beta)\} & = \E\left\{n^{-1/2} \sum_{i=1}^{n}I[Z_i \geq t_0] X_i \delta_i \frac{G(t_0)}{G(Z_i)}d_i(\beta)\right\}\\
&= n^{-1/2} \sum_{i=1}^{n} X_i \E \{d_i(\beta) | T_i \geq t_0\}.
\end{align*}
Let $\omega_{1i}^\ast$ be the line segment lying between $X_i^{\top} (\beta-\beta_0)$ and $X_i^{\top} (\beta-\beta_0)+\sigma_i t$. Then, 
\begin{align*}
& \E\{d_i(\beta) | T_i \geq t_0 \} = \int_{-\infty}^{\infty}d_i(\beta)g_{T_i - t_0} \{\epsilon_i(\beta)+ X_i^{\top}(\beta-\beta_0) | T_i \geq t_0 \} d\epsilon_i(\beta)\\
& = \sigma_i \int_{-\infty}^{\infty}\Phi(-\lvert t \rvert)\{2I[t>0]-1\} \left[g_{T_i - t_0}\{\sigma_i t + X_i^{\top}(\beta-\beta_0) \} + g_{T_i - t_0}^\prime\{\omega_i^\ast(t) \}\sigma_i t \right]dt
\end{align*}

It follows from Conditions C1 and C3 that $\sup_i g_{T_i - t_0} \{\sigma_i t + X_i^{\top} (\beta-\beta_0)\}<\infty$. Since $\int_{-\infty}^{\infty} \Phi(-\lvert t \rvert)\{2I[t>0]-1\}dt=0$, we have
\begin{equation*}
\int_{-\infty}^{\infty} \Phi(-\lvert t \rvert)\{2I[t>0]-1\}g_{T_i - t_0} \{X_i^{\star \top} (\beta-\beta_0)\}dt=0.
\end{equation*}
Again, by Condition C1, 
\begin{equation*}
\exists M > 0 \quad \textrm{such that} \quad \sup_i \lvert g_i^\prime \{\omega_i^\ast (t)\}\rvert < M.
\end{equation*} 
Thus, 
\begin{equation*}
\lvert E\{d_i(\beta)\} \rvert \leq \int_{-\infty}^{\infty} \lvert t \rvert \Phi(\lvert t \rvert) \lvert g_i^\prime \{\omega_i^\ast (t)\}\rvert dt \leq M \sigma_i^2 /2.
\end{equation*}
Note that $\sum_{i=1}^{n} \sigma_i^2 = \mbox{tr}(X H X^{\top}) = \mbox{tr}(H X^{\top}X)$ is bounded by $H=O(n^{-1})$ and Condition C2. Then, $\sum_{i=1}^{n} \lvert E\{d_i(\beta)\} \rvert \leq M \sum_{i=1}^{n}\sigma_i^2 /2$ is also bounded. Therefore,
\begin{equation}
\lVert \E\{D_n^{(1)}(\beta)\} \rVert \leq n^{-1/2} \sqrt{p} \sup\limits_{i,j} \lvert X_{ij} \rvert \sum_{i=1}^{n} \lvert \E\{d_i(\beta) | T_i \geq t_0 \} \rvert \to 0 \quad \text{as}\ n\to 0.
\end{equation}

By applying Condition C3, we have 
\begin{align*}
\Var\{D_n^{(1)}(\beta)\} & = \Var \Big\{n^{-1/2} \sum_{i=1}^{n} X_i X_i^{\top} I[Z_i \geq t_0]\delta_i \frac{G(t_0)}{G(Z_i)} d_i(\beta) \Big\}\\
& \leq n^{-1}\sum_{i=1}^{n} \frac{X_i X_i^{\top}}{c_0}\E\{d_i^2(\beta) | T_i \geq t_0\}.
\end{align*}
It follows from the arguments similar to evaluating $\E\{d_i(\beta) | T_i \geq t_0\}$ combining with Conditions C1 and C2, we have, as $n \to \infty$, $\lVert \E\{d_i^2(\beta) | T_i \geq t_0 \} \rVert \to 0$. This implies $\lVert \Var\{D_n^{(1)}(\beta)\} \rVert \to 0$. Then, by the Weak Law of Large Numbers, 
\begin{equation} \label{eq:12}
\lVert D_n^{(1)}(\beta) \rVert \xrightarrow{p} 0, \quad \text{as}\ n\to \infty.
\end{equation}
for $\beta$ in a compact neighborhood of $\beta_0$.

To show $\lVert D^{(2)}_n(\beta) \rVert \xrightarrow{p} 0$ as $n \to \infty$, we use the martingale representation of the Kaplan-Meier estimator \cite{fleming2011counting}. Specifically, $\hat{G}(t)$ can be represented as 
\begin{equation*}
\frac{\hat{G}(t) - G(t)}{G(t)} = -\sum_{i=1}^n \int_0^t \left\{ \frac{\hat{G}(u^-)}{G(u)}\right\}\frac{dM_{i}^C(u)}{Y(u)} 
\end{equation*}
where $M_i^C(u) = N_i^C(u)-\int_{0}^{t}Y_i(u)d\Lambda^C(s)$, $N_i^C(u)=(1-\delta_i)I[Z_i\leq u], \Lambda^C(u) = -\log\{G(u)\}$, $Y(u) = \sum_{i=1}^n Y_i(u)$, and $Y_i(u) = I[Z_i \geq u]$. By combining this with an application of the functional delta method and the uniform convergence result of $\hat{G}(\cdot)$ to $G(\cdot)$, we have 
\begin{align*}
D^{(2)}_{n}(\beta) & = n^{-1/2} \sum_{i=1}^{n}I[Z_i \geq t_0] X_i \delta_i n^{-1}\sum_{j=1}^n\left\{\frac{h_j(t_0)}{G(Z_i)} - \frac{h_j(Z_i)G(t_0)}{G^2(Z_i)}\right\} d_i(\beta) + o_p(1)\\
& = n^{-1/2} \sum_{j=1}^{n} \int_{t_{0}}^{\nu} \left\{n^{-1}\sum_{i=1}^n I[Z_i \geq t_0] X_i \delta_i Y_i(u) \frac{G(t_0)}{G(Z_i)}d_i(\beta)\right\} \frac{dM_{j}^C(u)}{y(u)}  + o_p(1)
\end{align*}
where 
\begin{equation*}
h_{j}(t) = G(t)\int_0^t \frac{dM_{j}^C(u)}{Y(u)} \mbox{ and }y(t) = \lim\limits_{n \to \infty} n^{-1} Y(t).
\end{equation*}
Using the arguments similar to those used to establish $\lVert D_n^{(1)}(\beta) \rVert \xrightarrow{p} 0$, as $n\to \infty$, it can be shown that $\E\left\{\left| Y_i(u)d_i(\beta) \right| \lvert \ T_i \geq t_0 \right\} = O(n^{-1/2})$. Thus, 
\begin{align*}
&\left\lVert \E\left\{n^{-1}\sum_{i=1}^n I[Z_i \geq t_0] X_i \delta_i Y_i(u) \frac{G(t_0)}{G(Z_i)}d_i(\beta)\right\} \right\rVert\\
& = \left\lVert n^{-1}\sum_{i=1}^n X_i \E\left\{Y_i(u)d_i(\beta) | T_i \geq t_0 \right\} \right\rVert \\
&\leq \sqrt{p} \sup\limits_{i,j} \lvert X_{ij} \rvert n^{-1} \sum_{i=1}^n \E \left\{\lvert Y_i(u)d_i(\beta) \rvert \ | \ T_i \geq t_0 \right\} \to 0.
\end{align*}
It then follows that, as $n \to \infty$
\begin{align*}
& \Bigg\lVert n^{-1}\sum_{i=1}^n I[Z_i \geq t_0] X_i \delta_i Y_i(u) \frac{G(t_0)}{G(Z_i)}d_i(\beta) \\
& - \E\left\{n^{-1}\sum_{i=1}^n I[Z_i \geq t_0] X_i \delta_i Y_i(u) \frac{G(t_0)}{G(Z_i)}d_i(\beta)\right\} \Bigg\rVert \xrightarrow{p} 0
\end{align*}
uniformly in $\beta$ for $\beta$ in the compact neighborhood of $\beta_0$. By applying the martingale central limit theorem and the Kolmogorov-Centsov Theorem \cite[p53]{karatzas1988brownian}, 
\begin{align*}
& n^{-1/2} \sum_{j=1}^{n} \frac{dM_{j}^C(u)}{y(u)} \text{converges weakly to a zero-mean Gaussian process}\\
& \text{with continuous sample paths.}
\end{align*}
By combining these results, it follows from Lemma 1 in \cite{lin2000fitting} that  
\begin{equation} \label{eq:13}
\Bigg\lVert n^{-1/2} \sum_{j=1}^{n} \int_{t_0}^{\nu} \bigg\{n^{-1}\sum_{i=1}^n I[Z_i \geq t_0] X_i \delta_i Y_i(u) \frac{G(t_0)}{G(Z_i)}d_i(\beta)\bigg\}\frac{dM_j^c(u)}{y(u)} \Bigg\rVert \xrightarrow{p} 0.
\end{equation}

By combining (\ref{eq:12}) and (\ref{eq:13}), we have
\begin{equation} \label{asymp:cnst:res}
\lVert {n^{1/2}} \{ \tilde{U}_n(\beta, \tau, H) - U_{t_0}(\beta, \tau) \} \rVert \xrightarrow{p} 0. 
\end{equation}
Note that both $\tilde{U}_{t_0}(\beta, \tau, H)$ and $U_{t_0}(\beta, \tau)$ are monotone functions, thus the point-wise convergence could be strengthened to uniform convergence \cite{shorack2009empirical}.

To establish the asymptotic equivalence of $n^{1/2}(\hat{\beta}_{IS} - \beta_0)$ and $n^{1/2}(\hat{\beta}_{NS} - \beta_0)$, it suffices to show that the following two convergence results hold: As $n \to \infty$,
\begin{align*}
&\mbox{(i) } \left\lVert \hat{A}(\beta_0, H) - A\right\rVert \to 0 \mbox{ and } \\
&\mbox{(ii) } \left\lVert n^{1/2}\left\{\tilde{U}_{t_0}(\beta_0, \tau, H) - U_{t_0}(\beta_0, \tau)\right\} \right\rVert \to 0.
\end{align*}
Note that Eq~\eqref{asymp:cnst:res} implies (ii). Thus, we prove (i). 
For any vectors $a, b \in R^p$,
\begin{align*}
\E \left[a^{\top} \hat{A}(\beta_0, H) b \right] & = a^{\top} \E \Bigg[ n^{-1}\sum_{i=1}^{n} I[Z_i > t_0] X_i X_i^{\top} \frac{\hat{G}(t_0) \delta_i}{\hat{G}(Z_i)} \phi\bigg(-\frac{\epsilon_i(\beta_0)}{\sigma_i}\bigg)\bigg(\frac{1}{\sigma_i}\bigg) \Bigg]b \\
&= a^{\top} \Bigg[n^{-1}\sum_{i=1}^{n} X_iX_i^{\top} \E\bigg\{\phi\bigg(-\frac{\epsilon_i(\beta_0)}{\sigma_i}\bigg) \bigg(\frac{1}{\sigma_i}\bigg)\bigg\lvert \ T_i > t_0, X_i \bigg\}\Bigg]b
\end{align*}
It follows from the variable transformation $t = \epsilon(\beta_0)/\sigma_i$ and the Taylor expansion at 0 that 
\begin{align*}
&\E \left\{\phi\bigg(-\frac{\epsilon_i(\beta_0)}{\sigma_i}\bigg)\bigg(\frac{1}{\sigma_i}\bigg) \bigg\lvert T_i > t_0, X_i \right\}\\
&= \int_{-\infty}^{\infty} \phi\bigg(-\frac{\epsilon_i(\beta_0)}{\sigma_i}\bigg)\bigg(\frac{1}{\sigma_i}\bigg)g_{T_i - t_0}(\epsilon_i)d\epsilon_i \\
&= \int_{-\infty}^{\infty} \phi\left(-t \right)g_{T_i - t_0}(0)dt + \sigma_i\int_{-\infty}^{\infty} t\phi\left(-t \right)g^{\prime}_{T_i - t_0}(\omega_{2i}^{\ast})dt
\end{align*}
where $\omega_{2i}^\ast$ is some value lying between 0 and $\sigma_i t$.

By Condition C1, we have 
\begin{equation*}
\sigma_i\int_{-\infty}^{\infty} t\phi\left(-t \right) g^{\prime}_{T_i - t_0}(\omega_{2i}^{\ast})dt \leq M\sigma_i\int_{-\infty}^{\infty} |t|\phi\left(-|t| \right)dt \to 0.
\end{equation*} 
Since $\int_{-\infty}^{\infty} \phi\left(-t \right)g_{T_i - t_0}(0)dt = 0$, we have $\E \left\{\phi\Big(-\frac{\epsilon_i(\beta_0)}{\sigma_i}\Big)\Big(\frac{1}{\sigma_i}\Big) \bigg\lvert T_i > t_0, X_i \right\} \to g_{T_i - t_0}(0)$ and, therefore,  
\begin{align}\label{asymp:slopeA:exp}
\lim_{n \to \infty} \E \left[a^{\top} \hat{A}(\beta_0, H) b \right] & = a^{\top} \left\{ \lim_{n \to \infty} n^{-1}\sum_{i=1}^n X_i X_i^{\top} g_{T_i - t_0}(0) \right\}b \\
&= a^{\top} A b. \nonumber
\end{align}
By Condition C1 and applying the arguments in \cite[p795, Appendix]{pang2012variance}, it can be shown that 
\begin{equation*}
\E \left[\left\{\phi\bigg(-\frac{\epsilon_i(\beta_0)}{\sigma_i}\bigg)\bigg(\frac{1}{\sigma_i}\bigg)\right\}^2 \bigg\lvert \ T_i > t_0, X_i \right] = O(n^{1/2}).
\end{equation*}

Then,
\begin{align}\label{asymp:slopeA:var}
&\Var \Bigg[a^{\top}\bigg\{n^{-1}\sum_{i=1}^{n} I[Z_i > t_0] X_iX_i^{\top} \frac{\hat{G}(t_0) \delta_i}{\hat{G}(Z_i)} \phi\bigg(-\frac{\epsilon_i(\beta_0)}{\sigma_i}\bigg)\bigg(\frac{1}{\sigma_i}\bigg)\bigg\} b\Bigg]\\
&\leq \ \frac{1}{n^2 c_0}\sum_{i=1}^n (a^{\top}X_i X_i^{\top})^2 \E\bigg[\bigg\{\phi\bigg(-\frac{\epsilon_i(\beta_0)}{\sigma_i}\bigg) \bigg(\frac{1}{\sigma_i}\bigg)\bigg\}^2 \bigg\lvert \ T_i > t_0, X_i \bigg] \to 0. \nonumber 
\end{align}

By combining the results in Eq~\eqref{asymp:slopeA:exp} and Eq~\eqref{asymp:slopeA:var}, we have $\left\lVert \hat{A}(\beta_0, H) - A\right\rVert \to 0$ as $n \to \infty$. This completes the proof of (i).

\section*{Appendix B. Supplementary material}
Supplementary material related to this article can be found online.

\section*{Acknowledgement}
This work was supported by the National Research Foundation of Korea(NRF) grant funded by the Korea government(MSIT) (2020R1A2C1A0101313911)

\section*{Data availability statement}
The code used in the simulation study is included in the supplementary information. For confidentiality reasons, the dental restoration dataset is available upon request.

\section*{Conflict of interest}
The authors declare that they have no conflict of interest.
 
\newpage
\bibliographystyle{spbasic}
\bibliography{residualQR}

\end{document}